\documentclass[twocolumn,rmp]{revtex4}

\usepackage{dcolumn,graphicx,amsmath,amssymb}

\usepackage{txfonts}

\begin{document}

\title{A network-based threshold model for the spreading of fads in
  society and markets}

\author{Andreas Gr\"{o}nlund}
\affiliation{Department of Physics, Ume{\aa} University, 90187
  Ume{\aa}, Sweden}

\author{Petter Holme}
\affiliation{Department of Physics, University of Michigan, Ann Arbor,
  MI 48109, U.S.A.}

\begin{abstract}
  We investigate the behavior of a threshold model for the spreading
  of fads and similar phenomena in society. The model is giving the
  fad dynamics and is intended to be confined to an underlying network
  structure. We investigate the whole parameter space of the fad
  dynamics on three types of network models. The dynamics we discover
  is rich and highly dependent on the underlying network
  structure. For some range of the parameter space, for all types of
  substrate networks, there are a great variety of sizes and
  life-lengths of the fads---what one see in real-world social and
  economical systems.
\end{abstract}

\maketitle

\section{Introduction and definitions}

\subsection{Background}

Society is an epitome of a complex system---at all levels it is driven by
non-equilibrium processes by heterogeneous sets of agents (or actors,
in sociologist speech). One of the more remarkable phenomena in our
everyday world is the presence of fads. How a certain mp3-player get
a substantial part of market, while other very similar products with
advertising budgets of a similar size, fade into obscurity? The same
question applies to everything from merchandise to sports, culture and
possibly even science. According to traditional economic
modeling~\cite{bik:fads} such phenomena are due to ``information
cascades'' in social networks. These can occur since the information
conveyed by the actions of a person's friends is more credible than
advertisement~\cite{bik:fads,arthur:contagion}, thus spreading from
one person to another can be an influential mechanism in a society with
mass-advertisements~\cite{bik:fads}. A popular class of models for the
spreading phenomena of this kind is so called threshold
models~\cite{rolfe:thrs,watts:fad,our:youth}. These, in general, serve
to model social and economic systems where the agents have a
resistance to change, but do change provided the motivation to do so
is big enough. Threshold models are attractive for physicists: They
are well-suited for the analytical and numerical techniques used by
statistical physicists.  Furthermore this kind of models are by their
nature defined on networks, so for the understanding of them one need
the theory of network structure~\cite{doromen:book,mejn:rev,ba:rev} (a
currently popular topic among interdisciplinary physicists). In this
report we study an extension of a threshold model proposed by
Watts~\cite{watts:fad} introduced as a model of the dynamics of youth
subcultures in Ref.~\cite{our:youth}. We perform a more detailed study
of its behavior on several underlying types of networks. We argue that
this model is applicable to a wide range of social spreading phenomena.

\subsection{Model}

Society and markets are non-equilibrium systems---if you see a twenty
year old picture of a downtown street scene, you would recognize
that it is not contemporary; if someone told you a ten year old mobile
phone was brand new, you would not be fooled. This direction of time
in society and markets is manifested through new things (commodity,
music, hobbies, beliefs, etc.)\ replacing old. In many cases a person
gets her, or his, motivation to change an old item to a new one by
friends, colleagues or other people in her, or his, social
surrounding. If many neighbors in a person's social network are doing,
or owning, something new, then that person is likely to follow the
neighbors' behavior. Another factor in this type of spreading is that
new things are more attractive than old, if this would not be the case
the above mentioned direction of time (manifested by new inventions
replacing old) would simply not exist. But there is also a resistance
to change that slows down the spreading of such habits.---for the sake
of convenience or old habits one may want to stick to the old rather
than changing to the new.  In summary, the three main principles of
the spread of fads are:
\begin{enumerate}
\item\label{p:majo} If the fraction of neighbors in the social network
  of a person currently adopting a certain fad is big enough, then
  that person will adopt that fad too.
\item\label{p:age} The attractiveness of a fad decreases with its age.
\item\label{p:res} There is a certain resistance to adopting a fad.
\end{enumerate}
(From now on we refer to all habits, merchandise, etc., as ``fads,''
regardless if they are large or small.)
Too keep the model simple, we need some further (little less
realistic) assumptions---for future studies of this problem these
assumptions can be relaxed.
\begin{enumerate}\setcounter{enumi}{3}
\item\label{p:slow} The underlying social network is changing much
  slower than the dynamics of the fads.---This means that we can keep the
  underlying network fixed and run our fad-simulations on top of it.
\item\label{p:1aat} An individual adopts one fad at a time.---In many
  cases we can assume that a fad in one area, say recreational sports or
  cell phones, is independent of fads in other areas, so that the
  model can be applied separately for each case.
\end{enumerate}
To combine points \ref{p:majo} and \ref{p:age} we assign, for
each individual $i$, a score $s_c(t,i)$ for every fad $c$. The score
is intended to represent the attractiveness of the fad to the
individual. If the score exceeds the threshold $T$, then the individual
adopts that fad. The score function we use is:
\begin{equation}\label{eq:s}
  s_c(t,i) = \frac{k}{k_i} n_i(c) \frac{t(c)-t(c_i)}{t-t(c_i)}
\end{equation}
where $k=2M/N$ is the average degree, $n_i(c)$ is the number of $i$'s
neighbors adopting fad $c$, $k_i$ is $i$'s degree and $t(c)$ is
the introduction time of fad $c$ and $c_i$ is $i$'s fad at time $t$. 
The factor $k/k_i$ rescales the score with respect to the degree of the
vertex. This enables us to use the same threshold for every vertex
while still fulfilling point \ref{p:majo}. The factor
$[t(c)-t(c_i)]/[t-t(c_i)]$ should be interpreted as the attractiveness of
a $c$ being proportional to the age difference between $c$ and the
vertex' current fad $c_i$ and inversely proportional to the age of $c$.

The dynamic model can thus be defined as:
\begin{enumerate}
\item Start with all vertices having the same fad. (The starting
  configuration is, in the limit of long run-times, negligible.) Let
  this initial fad have age zero at $t=0$.
\item\label{p:go} Go through the vertex set sequentially and, for
  each vertex $i$, calculate the score function $s_c(t,i)$.
\item Go through the vertex set sequentially once again.  If
  $s_c(t,i)>T$ change $i$'s current fad to $c$. If more than one
  fad exceeds the threshold then the one with highest score is
  adopted.
\item If the initial fad has vanished, save information about the fad
  configuration for statistics.
\item With a probability $R$ a new identity is assigned to a random
  vertex. So, on average, $NR$ fads are introduced per time step.
\item Increase the time counter (i.e.\ the time is measured in number
  of iterations) and go to point \ref{p:go} unless the simulation is
  finished.
\end{enumerate}
The only model parameter, apart from the network parameters and the
total time of the simulation, is the threshold $T$. We let the
simulation run for $50000$ time steps and $>10$ network realizations
(the precise number chosen to make errorbars sufficiently small---the
system is self-averaging so larger networks needs smaller averages).

\subsection{Networks}

We use three types of underlying model networks in our
simulations. The reason for this variety is twofold---first, the
structure of the type of social networks fads spread over is not
exactly known~\cite{our:seceder}; then, by comparing model networks
with well-known structural properties, one can conclude how the
different network structures influence the dynamical properties of the
network. The models we use are the Erd\H{o}s-R\'{e}nyi (ER) random
graphs~\cite{er:on}, the networked seceder model~\cite{our:seceder},
and a scale-free network model (SF) model~\cite{gronlund:sf}.

The Erd\H{o}s-R\'{e}nyi model is the simplest, most random, network
model. One starts from $N$ isolated vertices and add $M$ edges, one
by one, such that no multiple edges or self-edges are formed. These
networks are characterized by a very narrow distribution of the
degrees (Poissonian to be exact), a vanishing
clustering~\cite{wattsstrogatz} (density of triangles), and no
pronounced community structure~\cite{mejn:commu} (i.e.\ the feature
that the network can be clearly partitioned into subnetworks that are
densely connected within but sparsely interconnected). The ER model
lacks much of the structure (high clustering, pronounced community
structure, etc) that social networks are believed to
have~\cite{mejn:rev}. On the other hand, its lack of structure makes
it a good reference model to compare results from other models to. To
be well above the threshold for the emergence of a giant component
(which occurs when $M=N$)~\cite{janson} we set $M=2N$. Before applying
the dynamics we delete (the vanishingly small fraction of) vertices
and edges outside the giant component.

Our second network model is the networked seceder model. It is a model
designed to create networks with a strong community structure by
mimicking some features of social networking between individuals. For
its precise definition we refer to Ref.~\cite{our:seceder}. The
parameters of this model are the network sizes $N$ and $M$ and a
parameter $p$ controlling the strength of community structure---if
$p=1$ the network is of ER model character, if $p=0$ the network has
maximal community structure. Here we use $M=2N$ and $p=0{.}1$
throughout the  paper. Seceder model networks have (just like
acquaintance networks are believed to have) high clustering,
pronounced community structure, and a positive correlation between
degrees at either side of an edge~\cite{mejn:assmix}. The degree
distribution is exponentially decreasing (we note that some real-world
networks do have an exponential degree
distribution~\cite{our:seceder,amaral:classes}).

Both the ER and the seceder model have rather sharply peaked degree
distributions. As mentioned, it is not really clear what kinds of
degree distribution social networks have---probably different kinds of
social networks show different distributions. Since degree frequently
is power-law distributed we include a model generating networks with a
power-law distribution of degree. The method can in short be described
as a preferential attachment model~\cite{ba:model} where the network
grows both by the addition of stubs (a vertex and an edge with one end
attached to the vertex). The model has one parameter, $p$, that sets
the stub to edge addition ratio. A detailed presentation of the model
can be found in Ref.~\cite{gronlund:sf}. One starts with a connected
pair of vertices, and, at each time step, with probability $p$ add a
stub to the network. Then, with probability $1-p$, an additional edge
is added. Here we use $p=0{.}5$ to obtain the same density of edges,
$M=2N$, as for the other networks. In all steps edges are added
preferentially (i.e., the vertex to attach to is selected with a
probability proportional to the degree of the vertex). The degree
distribution at a given $p\in(0,1)$ and sufficiently large network is
a power-law $\sim k^{-\gamma}$ with an exponent
\begin{equation}\label{eq:gamma}
  \gamma = 2 + \frac{p}{2-p}
\end{equation}
The generated SF networks have a positive degree-degree correlation 
compared with a randomized version with the same degree 
distribution~\cite{gronlund:sf}, but the effects of correlations 
are not further investigated here.

\section{Simulation results}

\subsection{The time evolution of fads}

\begin{figure*}
  \resizebox*{0.7\linewidth}{!}{\includegraphics{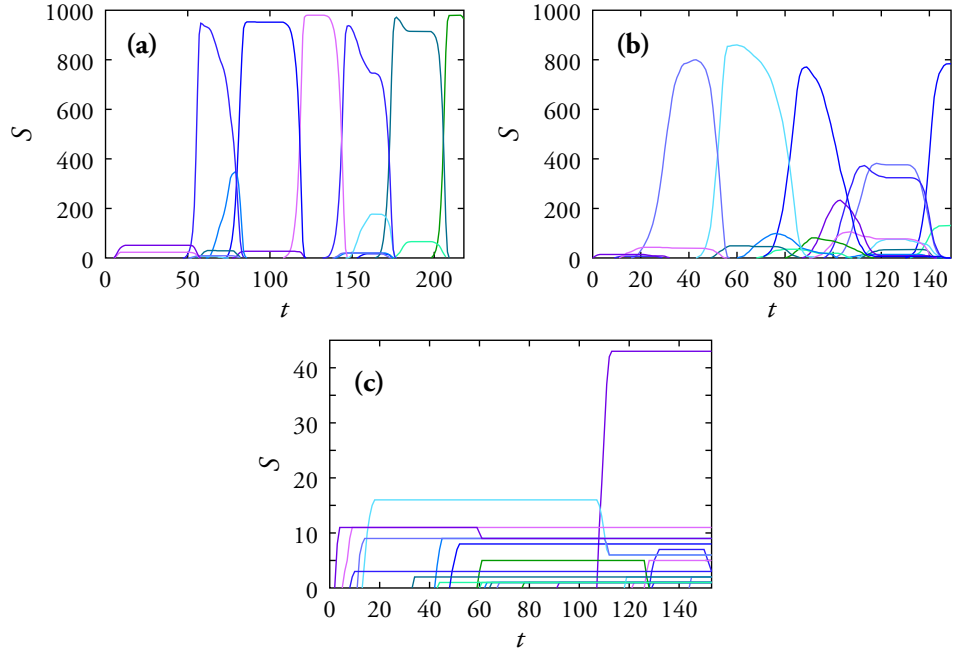}}
  \caption{ Examples of the time evolution of fads for three different
    underlying networks. (a) shows the initial time steps for the ER
    model networks. (b) and (c) are corresponding plots for the
    seceder and SF model networks. The sizes are $N=1000$,
    $M=2000$. The threshold is $T=0{.}7$.
  }
  \label{fig:series}
\end{figure*}

To get a first picture of the evolution of fads we plot the time
evolution of the size (number of adopters) $S$ in
Fig.~\ref{fig:series}. The ER and seceder model networks show a
rather similar behavior---for both these systems the effect of the
initial network seems to have disappeared within the interval
$t\lesssim 200$. For the SF networks the situation is radically
different---despite the similar threshold value ($T=0{.}7$), the
fads only spread to very limited surrounding. The reason for this is
the presence of hubs in the SF networks (i.e.\ vertices with a degree
far bigger than the average). The hubs have a larger influence on the
others, but are also less sensitive to new fads in their
surrounding. In the time evolution depicted in
Fig.~\ref{fig:series}(c) no fads manage to replace the initial fad
of a hub.

\begin{figure*}
  \resizebox*{0.85\linewidth}{!}{\includegraphics{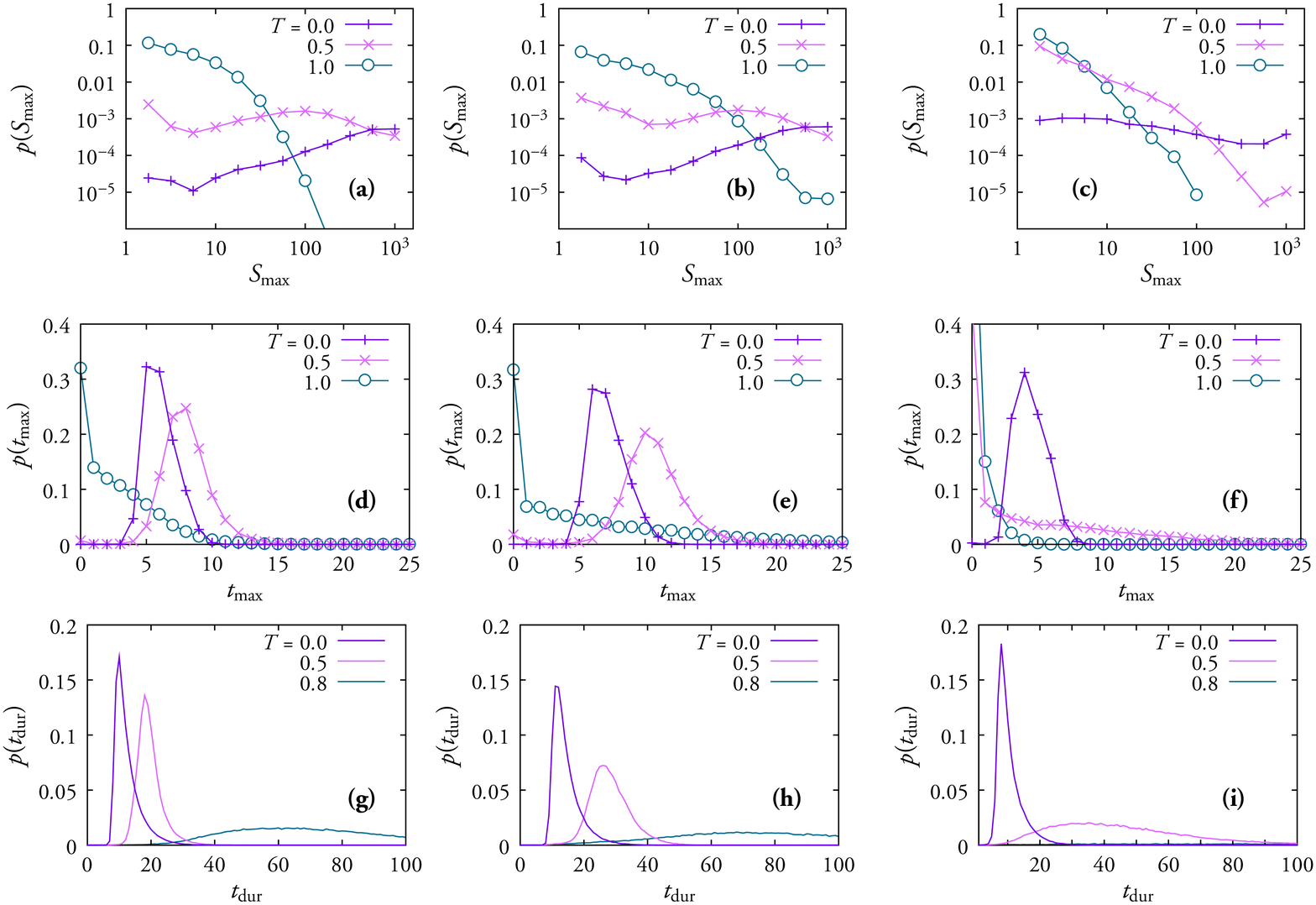}}
  \caption{ Distribution of some quantities for the different models:
  The distribution of the maximal fad size $S_\mathrm{max}$ for the ER
  (a), seceder (b) and SF (c) models. The distribution of time to the
  peak value of the fad $t_\mathrm{max}$ for ER (d), seceder (e) and
  SF (f) models. The distribution of life-lengths $t_\mathrm{dur}$ for
  ER (g), seceder (h) and SF (i) models. The network sizes are
  $N=1600$ and  $M=3200$. Errorbars
  are displayed if they are larger than the symbol size. Lines are
  guides for the eyes. }
  \label{fig:distri}
\end{figure*}

\begin{table}
  \caption{\label{tab:corr}
    Averages and correlations for the curves in
    Fig.~\ref{fig:distri}. The columns display the average largest size
    of the fads $\langle S_\mathrm{max} \rangle$, the average duration
    time of the fads $\langle t_\mathrm{dur} \rangle$, $\langle
    S_\mathrm{max} t_\mathrm{dur} \rangle$ which has a lower bound of
    $1/R=6400$ (attained for the case of rectangular fads , i.e.\ fads
    with no growth or recess stages of their life-time). We also measure
    the Pearson's correlation coefficient $r$ between $S_\mathrm{max}$
    and $t_\mathrm{dur}$. The numbers in parentheses indicate the
    standard error in the order of the last decimal place.}
  \begin{tabular}{cc|cccc}
    type & $T$ & $\langle S_\mathrm{max} \rangle$ & $\langle
    t_\mathrm{dur} \rangle$ & $\langle S_\mathrm{max} t_\mathrm{dur}
    \rangle$ & $r(S_\mathrm{max}, t_\mathrm{dur})$\\\hline ER & 0{.}0
    & 1097(2) & 12.53(9) & $1{.}46(1) \times 10^{5}$ & 0{.}524(4) \\ &
    0{.}5 & 637(2) & 19.54(6) & $1{.}282(4) \times 10^{5}$ &
    0{.}276(8) \\ & 1{.}0 & 8{.}8(3) & $2{.}1(1) \times 10^{3}$ &
    $2{.}7(1) \times 10^{5}$ & 0{.}129(7) \\\hline sec. & 0{.}0 &
    977(4) & 14.4(2) & $1{.}51(1) \times 10^{5}$ & 0{.}593(4)\\  &
    0{.}5  & 492(2) & 28.3(2) & $1{.}28(4) \times 10^{5}$ &
    0{.}42(1)\\ & 1{.}0 & 29(1) & 351(6) & $2{.}17(7) \times 10^{5}$ &
    0{.}32(2) \\\hline SF & 0{.}0 & 1189(1) & 10{.}55(5) & $1{.}34(7)
    \times 10^{4}$ & 0{.}416(5) \\ & 0{.}5 & 163(5) & 43(1) &
    $9{.}16(7) \times 10^{3}$ & 0{.}274(6)\\ & 1{.}0 & 3{.}1(1) &
    $2{.}84(7) \times 10^{3}$ & $1{.}57(6) \times 10^{4}$ &
    0{.}142(7)\\
  \end{tabular}
\end{table}

\subsection{Distribution of fad sizes and durations}

From Fig.~\ref{fig:series}(a) and (b) we see that $S$ really can grow
to very big fraction of the system size. But this does not mean that
there, in general, always can be fads of all sizes. Let the threshold
be fixed and finite and consider a network ensemble with fixed average
degree and a monotonically growing average distance between the vertices
(such as the three models considered---and, indeed, anything else would
be rather extreme). Since a fad can spread out from its origin one
edge at a time step, and since new fads can appear everywhere in the
network, there will almost surely be new fads to stop an old fad before they
reach a (big enough) fixed size $S'$ in the $N\rightarrow\infty$
limit. I.e., the probability that $S>S'$ goes to zero fast as
$S'\rightarrow\infty$. On the other hand, the large size limit does
not make much sense for social systems. The reason for this is that social
networks are of the small-world type~\cite{milg:1,watts:small2} with
extremely short average path lengths. Anything spreading from friends
to friends will only need the six degrees of
separation~\cite{watts:six} to reach an extension where the finite
size of humanity needs to be accounted for. This means our model
will not have phase where fads can grow without limit (like Watts'
model has~\cite{watts:fad}). If ``revival'' fads (retro fashion and
the like) are treated as new fads, this is not a problem---in the real
world there are simply no fads with unlimited staying
power. Even without fads that can grow boundlessly, the model can (of
course) show a broad spectrum of dynamic behavior. To investigate
this we start by plotting the probability distribution function of the
maximal number of adopters of a fad $S_\mathrm{max}$, for our three
network models and a number of threshold values (see
Fig.~\ref{fig:distri}(a), (b) and (c)). We see that the functional
form of $p(S_\mathrm{max})$ takes drastically different shapes of the
different parameter values. For $T=0$ the curves are almost
non-decreasing for all model networks. As mentioned above, the
monotonically increasing $p(S_\mathrm{max})$-curves are finite-size
effects (we will see this more clearly later). In
Fig.~\ref{fig:series}(d), (e) and (f) we plot the probability density
function of $t_\mathrm{max}$---the time it takes for a fad to reach
its maximum value. For low threshold values $p(t_\mathrm{max})$ has a
sharp peak. This observation---that fads reach their peak after a
characteristic time---is a possible test of the model (unfortunately we do
not know of such a data set). In  Fig.~\ref{fig:series}(g), (h) and (i)
we show the probability density function of the life-times of fads. We
note that the general shape of the $p(t_\mathrm{dur})$ curves is
rather similar to the $p(t_\mathrm{max})$-curves---the average and
the variance increase with $T$. However, the double peaks of the
$p(t_\mathrm{max})$-curves, for low threshold values, are now
gone. This means that the fads with an early peak does not go extinct
sooner than the fads of the second peak, they just do not enter a
stage of growth (i.e.\ they probably only consist of one or a few
vertices). The relation between the $p(t_\mathrm{max})$- and
$p(t_\mathrm{dur})$-curves can also tell us something about the
typical life span of a fad. On average, $NR$ fads are introduced per
time step, the average integrated time per fad is $1/R$:
\begin{equation}
  \frac{1}{R} = \frac{1}{n_\mathrm{tot}} \sum_{t=1}^{t_\mathrm{tot}}
  \sum_{i=1}^{n_\mathrm{tot}} S(i,t) < \frac{1}{n_\mathrm{tot}}
  \sum_{i=1}^{n_\mathrm{tot}} S_\mathrm{max}(i)t_\mathrm{dur}(i)
  \langle S_\mathrm{max}t_\mathrm{dur}\rangle
\end{equation}
If $\langle S_\mathrm{max}t_\mathrm{dur}\rangle$ is close to
$1/R$ the shape of a fad (in a $S(t)$-plot) will be near
rectangular. In Tab.~\ref{tab:corr} we list values of $\langle
S_\mathrm{max}t_\mathrm{dur}\rangle$ (for the curves of
Fig.~\ref{fig:distri}) along with values of $S_\mathrm{max}$,
$t_\mathrm{dur}$ and the correlation between the two latter
quantities. What we find is that the SF model network have $\langle
S_\mathrm{max}t_\mathrm{dur}\rangle$-values quite close to $1/R$
($0{.}5$-$2{.}4$ times larger), thus for these networks it may be
relevant to divide the life-time of a fad into a growth stage, a
quasi-stationary stage and stage of decline. The other networks have
$\langle S_\mathrm{max}t_\mathrm{dur}\rangle$-values far above
$1/R=6400$, we can thus conclude that fads in these network have a
much slower growth or decline than fads in the SF model networks.

\subsection{What determines the size of the fad?}

\begin{figure*}
  \resizebox*{0.7\linewidth}{!}{\includegraphics{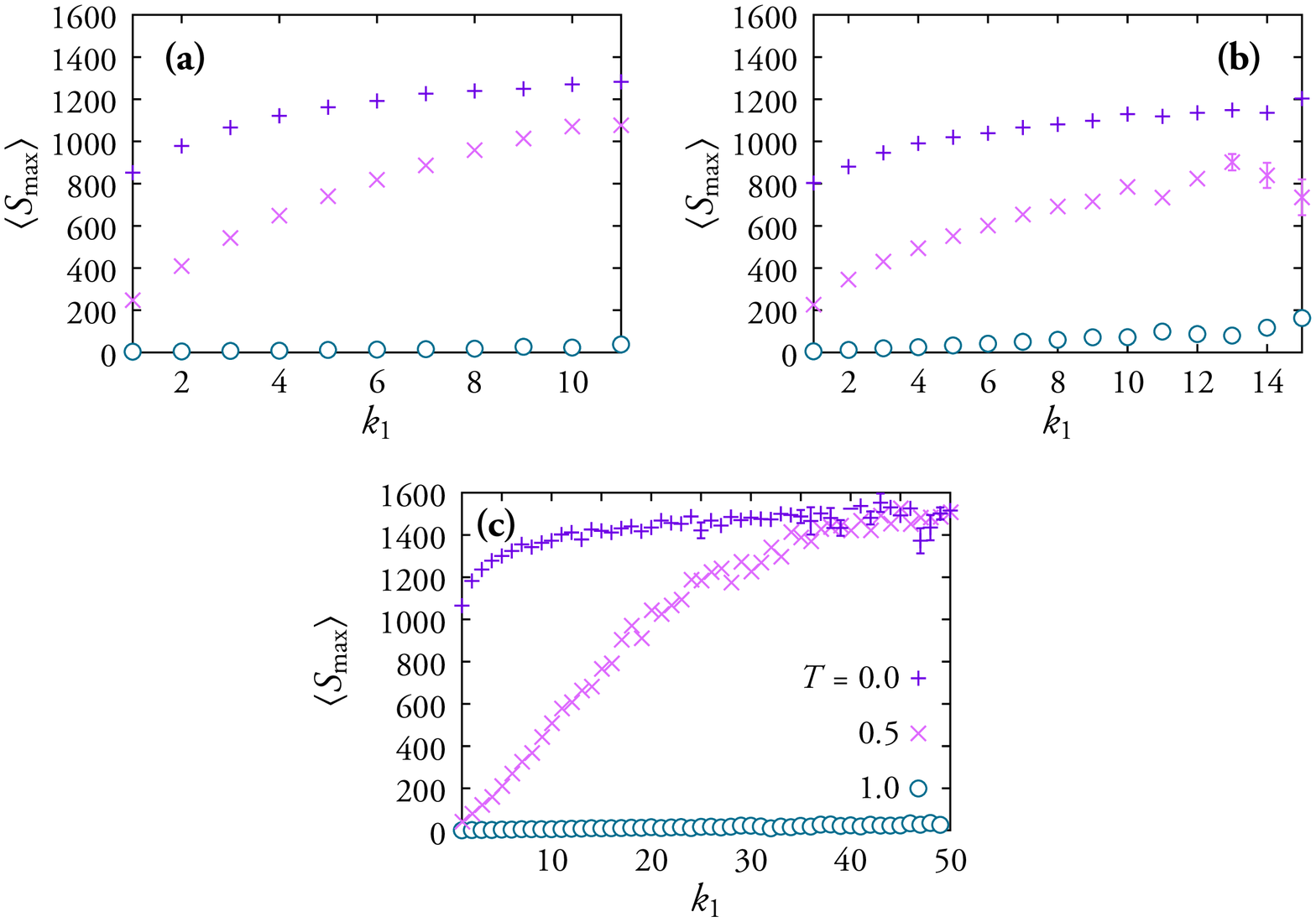}}
  \caption{
    The maximal size $S_\mathrm{max}$ of a fad as a function of
    the degree of its first adopter. The network sizes are $N=1600$
    and $M=3200$. The panels show (a) ER, (b) seceder and (c) SF model
    networks. }
  \label{fig:n_deg}
\end{figure*}

\begin{figure*}
  \resizebox*{0.7\linewidth}{!}{\includegraphics{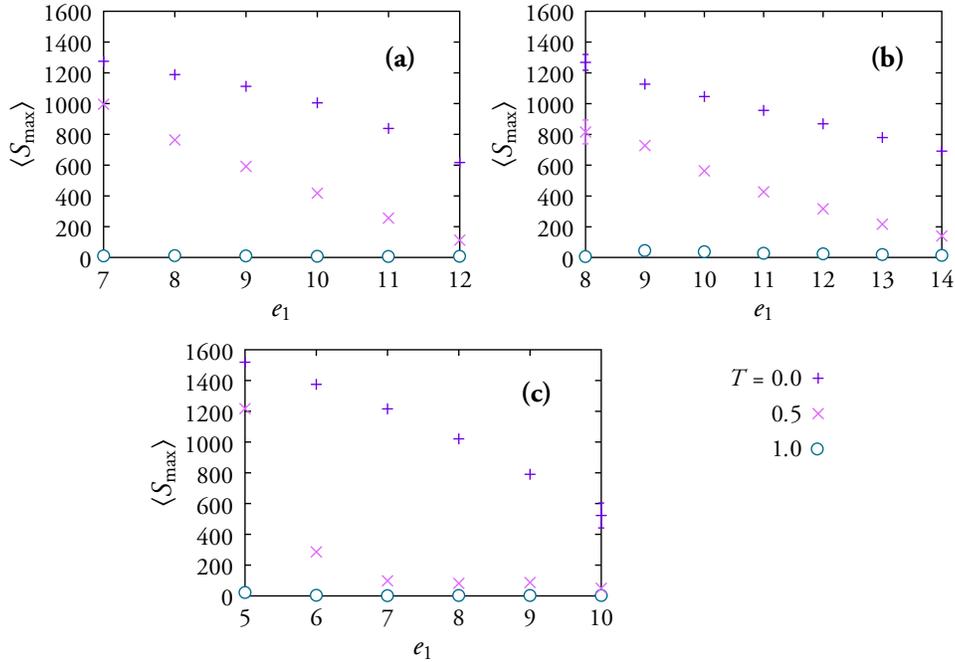}}
  \caption{
    The maximal size $S_\mathrm{max}$ of a fad as a function of
    the eccentricity $e_1$ of its first adopter. (a) shows the ER, (b)
    shows the seceder and (c) shows the SF model networks.
  }
  \label{fig:n_ecc}
\end{figure*}

The early time-evolution of a new fad depends on the age and
configuration of fads in the surrounding of the first
adopter. Another factor is the network characteristics of the first
adopter. For example, if the first adopter has a high degree, there are more
people the fad can spread to, and thus the chances for it to
spread will increase. To test this, we plot the average maximal size
$S_\mathrm{max}$ conditioned on the degree of the first adopter in
Fig.~\ref{fig:n_deg}. As expected for all network types and threshold
values $S_\mathrm{max}$ is (within the errorbars) strictly increasing
with the degree of the first adopter $k_1$. The broad degree
distribution of the SF-model networks is also strengthening this
effect. The increase seems to be most dramatic for low-degree vertices
and intermediate $T$-values---for the $k_1=3$ vertices of SF model
networks $S_\mathrm{max}$ increase over 15 times when $T$
decreases from $0{.}5$ to $0$.

Another network property than can influence the size of the fad is the
centrality of the first adopter. If a fad starts at a peripheral
vertex, it would be old already at the time it reaches the more
central regions. As seen in Fig.~\ref{fig:n_ecc} this is indeed true
for almost all network models and threshold values (the one exception
is the $T=1$ curve in Fig.~\ref{fig:n_ecc}(b) where the $e_1=8$ point
lies below the $e_1=9$ point). The effect is (just as for the degrees of
the first adopter) strongest for the SF networks with
$T=0{.}5$. Networks with a power-law degree distribution are known to
have a very compact core within which the average path lengths scale as
$\log\log N$ (to be compared with the $\log N / \log\log N$ scaling in
the graph as a whole)~\cite{chung_lu:pnas,cohen:ultra}. It is thus not
a surprise that the fads starting in the core ($e_1=5$ in
Fig.~\ref{fig:n_ecc}(c)) are more likely to spread to a large
population than the peripheral vertices.

\subsection{Finite-size scaling of the fad sizes}

\begin{figure*}
  \resizebox*{0.7\linewidth}{!}{\includegraphics{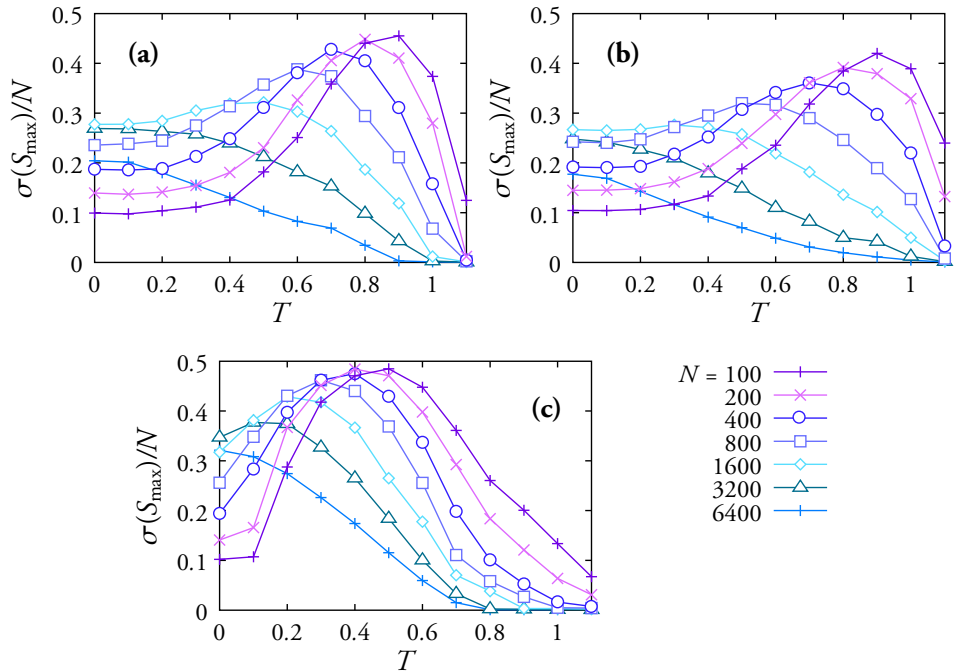}}
  \caption{
    Finite size scaling plots of the standard deviation of the maximal
    fad sizes divided by the system size. (a) shows the results for
    the ER model, (b) is the corresponding plot for the seceder model
    and (c) shows the curves for the SF model. The density of edges is
    constant $M=2N$. Lines are guides for the eyes. Errorbars are
    smaller than the symbol size.
  }
  \label{fig:fss}
\end{figure*}

As discussed above we do not expect a phase where the fads can grow
boundlessly. To investigate this further we plot the standard
deviation of the maximal fad size divided by the system size $N$
(Fig.~\ref{fig:fss}). In a situation where variance of the maximal
sizes of the fads does not diverge, this quantity will tend to zero as
the system size increases. As expected, this is exactly what we observe 
for all networks and threshold values. For networks with a small
diameter, and thresholds that allow high growth rates of the fads, the
finite system size will limit the growth of a significant fraction
of the initiated fads. Therefore, for small systems sizes and
threshold values, the variance appears to diverge as $N$ grows. For a
sufficiently large network though (in which one observes the maximum
sizes of the fads to be significantly smaller than the size of the
network); the growth of a fads will, in general, not be limited by the
boundary. Thus the variance will in this case not be bounded by the
finite size of the system, but rather be bounded by the appearance of
new (and thus more attractive) fads at the boundary of the fad. In
this situation the maximal size of a fad highly depends its possible
growth rate while being young, and thus on the network structure. This
implies that scaling up the network without altering its topological
characteristics, will not produce larger fluctuations. Specifically,
if we consider the fluctuations at the threshold $T=0$, we see that
the relative size of the fluctuations grows with the system size until
a certain $N$ is reached, and from there on it does not increase with
the system size. For the biggest systems simulated, $N=6400$, the
largest fluctuations at $T=0$ are found in the SF networks because of
the potentially much faster growth rate of a fad here than in the
other networks. 

The peaked shape of the $\sigma(S_\mathrm{max})/N$ vs.\ $N$ curves can
be explained by two competing mechanisms governing the variance of the
fads; if $T$ is small newer fads will spread to vertices currently
occupied by older fads until they get old and unattractive or replaced
by new fads, and it is reasonable to believe $S_\mathrm{max}$ will be
sharply peaked around its average in this case; if $T$ is large, most
fads will die out as soon they are born, some fads may spread to a
large population but not many enough to make the variance large. In
the real world we expect the fads to have a rather broad, but
decreasing, distribution of maximal sizes~\cite{watts:fad,our:youth},
a situation resembling intermediate $T$-values.

\section{Summary and conclusions}

The spreading of fads is a peculiar and poorly understood phenomenon
in social and economic systems. In this paper we present a thorough
investigation of a dynamical model for the spreading of fads put on
three types of underlying complex network models: Erd\H{o}s-R\'{e}nyi
random graphs, the networked seceder model and a model generating
networks with power-law distributed degrees. The reason to use several
underlying network models is that the network structure of social
networks in general (and the kind of social network fads spread over
in particular) is in several aspects unknown. The reason we include
the Erd\H{o}s-R\'{e}nyi model is that it is the simplest, most
well-studied and most random network model. The networked seceder
model captures many features---assortative mixing, high clustering and
community structure---that social networks are believed to
have. Studies of some types of social networks (sexual
networks~\cite{liljeros:sex} and networks of electronic
communication~\cite{aiello,pok,bornholdt:email}) report fat-tailed
distributions of degree, something the Erd\H{o}s-R\'{e}nyi and seceder
model networks lack. For this reason we also include the model
producing networks with a power-law degree distribution.

The fad dynamics is based on five assumptions about the individual's
responses to his/her social surrounding. In brevity, a person is only
adopting one fad at a time, and (s)he is willing to adopt a new fad
only if its attractiveness exceeds a certain threshold value. The
attractiveness of a fad increases with the number of network neighbors
that are currently following that particular fad, and decreases with
the age of the fad. For small threshold values, the life-length of a
fad is rather sharply distributed whereas the maximal size can take a
broad range of values. For high thresholds, the probability
distribution of the life-time of a fad decreases slowly (i.e.\ some
fads live a very long time, but most fads die as soon as they appear),
and the distribution of maximal sizes is decaying rather fast. In the
intermediate regime there are fads of all kinds of sizes and life
lengths. While this general picture is true for all three underlying
network models other features are different between the models: The
shape of the time-evolution (i.e.\ the functional shape of the size of
the fad $S$ vs.\ time $t$) differs---the SF model has fads with
distinct stages of growth and decline, whereas the ER and seceder
models have more complex time evolutions (being much smaller than
their maximal value most of the time). Furthermore, we investigate how 
the size of the fad depends on the network characteristics of the
first adopter. We find that a fad is more likely to be large if the
first adopter has a high degree or a low eccentricity.

Our model captures some known features of fad-sensitive social and
economic systems, like a wide-distribution of fad sizes and duration
times~\cite{watts:fad,our:youth}; and other features that seem very
plausible, like that the largest fads typically start at socially
well-connected and central persons. This field would however benefit
substantially from quantitative data, both regarding how individuals
respond to their social surrounding (in terms of their fads) and the
time evolution of the fads themselves. We note that, in the respect,
the study of fad-dynamics lags behind related fields like the study of
voluntary organizations~\cite{liljeros:phd}.

\end{document}